\documentclass[conference]{IEEEtran}
\usepackage{latexsym}
\usepackage{graphicx}
\usepackage{amsfonts,amssymb,amsmath}
\usepackage{mathtools, cuted}
\usepackage{hyperref}

\usepackage[T1]{fontenc}
\usepackage{cite}
\usepackage{subcaption}
\usepackage{comment}

\usepackage{amsthm}
\usepackage{overpic}
\usepackage{steinmetz}
\usepackage{array}
\usepackage{url}
\usepackage{color}

\usepackage{algorithm}
\usepackage{algorithmic}

\theoremstyle{plain}

\newtheorem{definition}{Definition}

\newtheorem{proposition}{Proposition}

\usepackage{flushend}

\newcommand{\vect}[1]{\mathbf{#1}}

\def\Htran{\mbox{\tiny $\mathrm{H}$}}
\def\Ttran{\mbox{\tiny $\mathrm{T}$}}
\IEEEoverridecommandlockouts

\begin{document}

\title{Broad Beam Reflection for RIS-Assisted \\ MIMO Systems with Planar Arrays}

\author{Parisa Ramezani\textsuperscript{1}, Maksym A. Girnyk\textsuperscript{2}, and \thanks{This work was supported by the Swedish Research Council, 2019-05068.} Emil Bj\"{o}rnson\textsuperscript{1}\\
\IEEEauthorblockA{\textit{\textsuperscript{1} Department of Computer Science, KTH Royal Institute of Technology, Stockholm, Sweden} \\ \textit{\textsuperscript{2}Ericsson Networks, Stockholm, Sweden}\\ Email: \{parram, emilbjo\}@kth.se, max.girnyk@ericsson.com}}

\maketitle

\begin{abstract}
  While reconfigurable intelligent surface (RIS)-aided user-specific beamforming has been vastly investigated, the aspect of utilizing RISs for assisting cell-specific transmission has been largely unattended. Aiming to fill this gap, we study a downlink broadcasting scenario where a base station (BS) sends a cell-specific signal to all the users located in a wide angular area with the assistance of a dual-polarized RIS. We utilize the polarization degree of freedom offered by this type of RIS and design the phase configurations in the two polarizations in such a way that the RIS can radiate a broad beam, thereby uniformly covering all azimuth and elevation angles where the users might reside. Specifically, the per-polarization configuration matrices are designed in such a way that the total power-domain array factor becomes spatially flat over all observation angles implying that the RIS can preserve the broad radiation pattern of a single element while boosting its gain proportionally to its aperture size. We validate the mathematical analyses via numerical simulations.   
\end{abstract}
\begin{IEEEkeywords}
Dual-polarized beamforming, Golay complementary pairs, power-domain array factor.
\end{IEEEkeywords}

\section{Introduction}
In recent years, we have witnessed an upsurge of interest in the development of reconfigurable intelligent surface (RIS)-assisted communication systems. An RIS is an auxiliary network entity engineered to control the propagation of electromagnetic waves intelligently \cite{Wu2020a}. Despite the extensive research conducted in this area, some fundamental problems have remained untouched. Particularly, the majority of prior works have considered user-specific RIS-aided communication, where the phase shifts of RIS elements are designed in a way to form a narrow beam towards one (or a few) specific user(s) \cite{Guo2020, Dong2020}. In many practical situations, however, many users that spread over a wide angular sector and possibly at unknown locations must be simultaneously served. An example of such a scenario is when the base station (BS) broadcasts cell-specific reference and synchronization signals. 
Therefore, if an RIS is deployed to assist the communication between the BS and the users in such cases, a broad beam needs to be radiated from the surface to uniformly cover all angular directions of interest. Furthermore, broad beams can be used for small-packet transmission, low-latency transmission, and in high mobility scenarios \cite{ericsson2017forming}. 

The beamwidth of a focused transmission is inversely proportional to the aperture size; therefore, a larger RIS produces a narrower beam \cite{Bjornson2020a}. This makes the broad beam design for RIS-aided communication challenging since a practical RIS is expected to have a large aperture. Recently, beam-broadening methods have been proposed in \cite{He2023BroadCoverage} and \cite{AlHajj2023} for partially widening the beam reflected by the RIS. In particular, reference \cite{He2023BroadCoverage} aims to achieve a quasi-static broad coverage by minimizing the difference between
a predefined pattern and the RIS-reflected power pattern, while the authors in \cite{AlHajj2023} design a codebook of wide beams with some constraints on the side-lobe level. The proposed approaches, however, fail to produce a perfectly broad beam even in a limited angular interval. As will be elaborated in detail later in the paper, a perfectly broad beam is a beam whose power-domain array factor is constant over all angular directions of interest.

Although designing broad beams for large uni-polarized arrays is problematic, it is possible to produce perfectly broad radiation patterns using dual-polarized arrays \cite{Max2021,Li2021Golay}. The broadness of the beam is achieved by combining per-polarization radiated beams designed to complement each other.
In our recent work in \cite{ParisaBroadBeamLetter}, we utilized the polarization degree of freedom of a ULA-type RIS and designed the per-polarization RIS configuration vectors to obtain a uniformly broad reflected beam. This work extends our previous work by considering the more practical uniform planar array (UPA)-type RIS and designs the phase configurations of the dual-polarized RIS such that the power-domain array factor of the surface becomes spatially flat, thus letting the RIS radiate a broad beam. Specifically, the radiated beam needs to uniformly cover all target azimuth and elevation angles which makes the design of RIS phase configurations more challenging compared to the ULA case in \cite{ParisaBroadBeamLetter} where only azimuth angles were considered. We show that a perfectly uniform radiated beam is only achievable if the auto-correlation function (ACF) of the RIS configuration matrices add up to a Kronecker delta function, a property possessed by the so-called \textit{Golay complementary array pairs} \cite{Golay1951, jedwab2007golay}. We validate the theoretical analyses through simulations. 

\section{System Model}
\begin{figure}[t]
    \centering
    \includegraphics[width = 0.65\columnwidth]{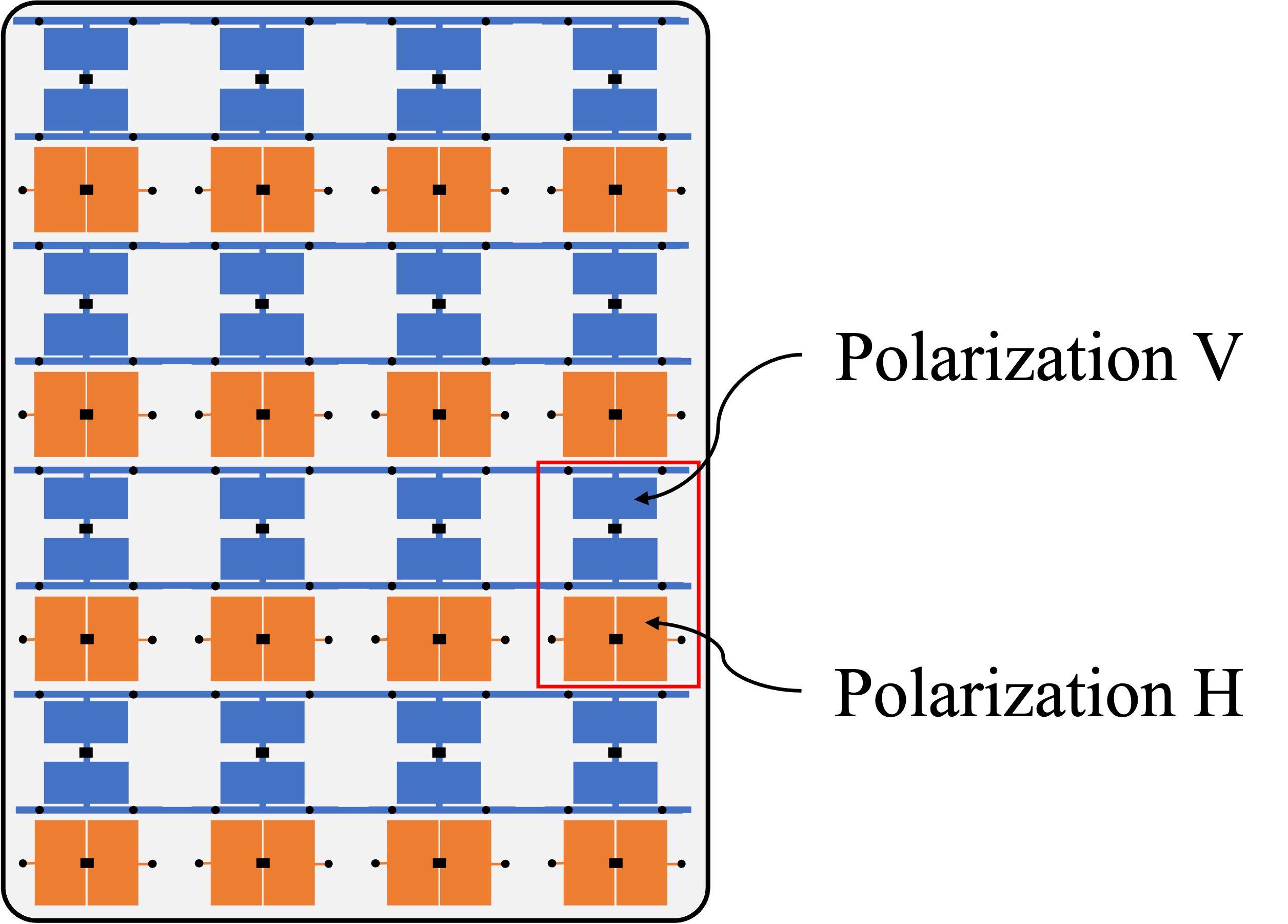}
   \caption{A dual-polarized RIS consisting of meta-atoms with two orthogonal separately controllable polarizations \cite{Chen2021,ke2021linear}.} 
    \label{fig:dp-RIS}
\end{figure}

We consider an RIS-assisted system with a BS communicating with a number of users via the RIS. In particular, we study a broadcast communication scenario where the BS intends to transmit a common signal to all the users who reside in a wide angular sector, and possibly at unknown locations.
The direct links between the BS and users are assumed to be blocked and communication can only take place through the RIS.
The BS is equipped with $M$ dual-polarized antennas and the users have one dual-polarized antenna each. The RIS is assumed to have $2N$ elements, out of which, $N$ elements have $\mathrm{H}$ polarization and the other $N$ elements have $\mathrm{V}$ polarization.\footnote{$\mathrm{H}$ and $\mathrm{V}$ refer to horizontal and vertical polarizations, but the results of this paper hold for any pair of orthogonal polarizations.} The elements are arranged in a way that the polarization changes between different RIS rows, as depicted in Fig.~\ref{fig:dp-RIS}.

Suppose $s$ is the signal transmitted by the BS. Assuming line-of-sight (LOS) channels between the BS and the RIS and the RIS and the users,\footnote{Based on a recent study in \cite{Emil2022},  LOS channel between the transmitter and RIS and between the RIS and receiver is the most practical scenario. Therefore, we aim to achieve uniform array gain in all LOS directions.} the received signal in polarization $\mathrm{p} \in \{\mathrm{H,V}\}$  at a potential user in the azimuth angle $\varphi$ and elevation angle $\theta$ from the RIS can be expressed as 
\begin{equation}
\begin{aligned}
\label{eq:received_signal}
   r_{\mathrm{p}} =&\; \sqrt{M P_{\mathrm{T}} \beta_2 \beta_1 G_{\mathrm{B,}0}(\vartheta)G_{\mathrm{R,}0}(\Tilde{\varphi},\Tilde{\theta})G_{\mathrm{R,}0}(\varphi,\theta)}\\ &\times \boldsymbol{\phi}_{\mathrm{p}}^{\Ttran} \left( \vect{a}_{\mathrm{p}} (\varphi,\theta) \odot \vect{a}_{\mathrm{p}} (\tilde{\varphi},\tilde{\theta})  \right)s + n_{\mathrm{p}}, 
   \end{aligned}
\end{equation}where $P_{\mathrm{T}}$ is the BS transmit power, $\beta_1$ and $\beta_2$ denote the path-loss from the BS to an RIS element and from an RIS element to the users, and $G_{\mathrm{R},0}(\cdot)$ and  $G_{\mathrm{B},0}(\cdot)$ represent the radiation power pattern of one RIS element and one BS antenna, respectively. Moreover,  
$\vect{a}_{\mathrm{p}}(\cdot)$ denotes the RIS array response vector in polarization $\mathrm{p}$, and $\boldsymbol{\phi}_{\mathrm{p}} = \left[e^{j\phi_{\mathrm{p},1}},\ldots,e^{j\phi_{\mathrm{p},N}}\right]^{\Ttran}$ is the RIS phase shift configuration vector with $\phi_{\mathrm{p},n}$ denoting the phase shift applied by the $n$th RIS element in polarization $\mathrm{p}$ to the incident signal.  $\vartheta$ is the angle of departure (AoD) from the BS, and $\Tilde{\varphi}$ and $\Tilde{\theta}$ are the azimuth and elevation angle of arrival (AoA) at the RIS. Finally, $n_{\mathrm{p}}$ is the noise at the user.

The array response vectors for the RIS is given by 
\begin{equation}
    \vect{a}_{\mathrm{p}}(\varphi,\theta)= \vect{a}_{\mathrm{p}}^z (\theta) \otimes \vect{a}_{\mathrm{p}}^y(\varphi,\theta),~~\mathrm{p}\in\{\mathrm{H,V}\},
\end{equation}with $\vect{a}_{\mathrm{p}}^y(\varphi, \theta)$ and $\vect{a}_{\mathrm{p}}^z (\theta)$ denoting the RIS response vectors in $y$ and $z$ dimensions. The number of elements in each row and column of the RIS is respectively denoted by $N_y$ and $N_z$ such that $2N=N_y N_z$, where $N_z$ is assumed to be an even number to have the same number of elements in both polarizations, while $N_y$ can be even or odd. The array response vectors are given by \cite{massivemimobook}
\begin{align}
   \vect{a}_{\mathrm{H}}^y (\varphi,\theta)&=  \big[1, e^{-j \psi_y },\ldots,e^{-j(N_y -1)\psi_y} \big]^{\Ttran}, \\
   \vect{a}_{\mathrm{H}}^z (\theta) &=  \Big[1, e^{-j2\psi_z},\ldots,e^{-j2(\widetilde{N}_z -1)\psi_z} \Big]^{\Ttran},
\end{align}for the $\mathrm{H}$ polarization, where 
\begin{equation}
\label{eq:psi}
    \psi_y = \frac{2\pi}{\lambda} \Delta_y \sin(\varphi)\cos(\theta),~~\psi_z =\frac{2\pi}{\lambda} \Delta_z \sin(\theta) 
\end{equation} are the relative phase shifts in the $y$ and $z$ dimensions and $\widetilde{N}_z = N_z/2$. $\Delta_y$ and $\Delta_z$ denote the inter-element spacings between adjacent elements in respective dimensions, and $\lambda$ is the wavelength of the transmitted signal. For polarization $\mathrm{V}$, we have $\vect{a}_{\mathrm{V}}^y(\varphi,\theta) = \vect{a}_{\mathrm{H}}^y (\varphi, \theta)$ and $\vect{a}_{\mathrm{V}}^z (\theta)= e^{-j\psi_z} \vect{a}_{\mathrm{H}}^z(\theta)$ due to the vertical shift by one row, which results in 
\begin{equation}
\vect{a}_{\mathrm{V}}(\varphi,\theta) = e^{-j\psi_z} \vect{a}_{\mathrm{H}} (\varphi,\theta).
\end{equation} 

After applying maximum ratio combining of the signal received over two polarizations, the prospective user observes the total radiation power pattern 
\begin{equation}
\label{eq:radiation_pattern}
    G(\varphi,\theta) = A(\varphi,\theta)G_{\mathrm{R},0}(\Tilde{\varphi},\Tilde{\theta})G_{\mathrm{R},0}(\varphi,\theta),
\end{equation}from the RIS where the term 
\begin{equation}
\label{eq:PDAF}
A(\varphi,\theta) = \left|\boldsymbol{\phi}_{\mathrm{H}}^{\Ttran}\big(\hat{\vect{a}}_{\mathrm{H}}(\varphi,\theta)\big)\right|^2 + \left|\boldsymbol{\phi}_{\mathrm{V}}^{\Ttran}\big( \hat{\vect{a}}_{\mathrm{V}}(\varphi,\theta)\big) \right|^2
\end{equation} is referred to as the \textit{power-domain array factor} with $\hat{\vect{a}}_{\mathrm{p}}(\varphi,\theta) = \vect{a}_{\mathrm{p}}(\varphi,\theta) \odot \vect{a}_{\mathrm{p}}(\tilde{\varphi},\tilde{\theta}),~\mathrm{p}\in \{\mathrm{H,V}\}$.  The aim of this paper is to design the RIS configurations in $\mathrm{H}$ and $\mathrm{V}$ polarizations such that the beam re-radiated from the RIS covers all angular directions, which is achieved when the array factor is constant regardless of where the user is. The details of the proposed design will be unfolded in the next section.

\section{Broad Beam Design}
A perfectly broad beam is referred to as a beam whose power-domain array factor is spatially flat over all possible observation angles $(\varphi,\theta)$, i.e.,
\begin{equation}
    \label{eq:PDAF_constant}
    A(\varphi,\theta) = c,~~\varphi \in \left[-\frac{\pi}{2},\frac{\pi}{2} \right],~\theta \in \left[-\frac{\pi}{2},\frac{\pi}{2} \right].
\end{equation}Note that the proposed design is not affected by the AoAs, and $\Tilde{\varphi}$ and $\Tilde{\theta}$ can take any value, as will be further elaborated in the following. 

\subsection{Golay Complementary Pairs}
To design the RIS configurations for the $\mathrm{H}$ and $\mathrm{V}$ polarizations, we first describe the concept of Golay complementary sequence pairs, which was introduced by Golay in \cite{Golay1951}.

\begin{definition}[Golay complementary sequence pair]
\label{def:Golay-pair}
    Unimodular sequences $\vect{u} \in \mathbb{C}^N$ and $\vect{w} \in \mathbb{C}^N$ form a Golay complementary sequence pair if 
    \begin{align}
    \label{eqn:conditionGolay}
    	R_{\vect{u}}[\tau] + R_{\vect{w}}[\tau] &= 2N \delta [\tau],
    \end{align}where $R_{\vect{u}} [\tau]$ indicates the ACF of $\vect{u}$ and is given by 
    \begin{small}
    \begin{align}
    \label{eq:ACF}
    	R_{\vect{u}}[\tau] = \left\{ 
    	\begin{aligned} 
    	&\sum\limits_{n=1}^{N-\tau} [\vect{u}]_n [\vect{u}]_{n+\tau}^*, 
    	 & & \tau=0,\ldots,N-1,\\
    	&\sum\limits_{n=1}^{N+\tau} [\vect{u}]_{n-\tau} [\vect{u}]_{n}^*, 
    	 & &\tau = -N+1,\ldots,-1,\\
    	& \qquad  0,
    	 & & \mathrm{otherwise}.
    	\end{aligned} \right. 
    \end{align}
      \end{small}
\end{definition} 
Now, let $S_{\vect{u}}(f)$ be the power spectral density (PSD) of the sequence $\vect{u}$. According to the Wiener-Khinchin theorem, the PSD and ACF are Fourier transform
pairs \cite{Wiener1930}, i.e., 
\begin{equation}
   S_{\vect{u}}(f) = \sum_{\tau = -N+1}^{N-1}R_{\vect{u}}[\tau]\, e^{-j2\pi f \tau}.
\end{equation}For a Golay complementary pair $(\vect{u},\vect{w})$, we have 
\begin{equation}
\begin{aligned}
   S_{\vect{u}}(f) + S_{\vect{w}}(f) &= \sum_{\tau = -N+1}^{N-1} \left( R_{\vect{u}}[\tau] + R_{\vect{w}}[\tau]\right)\, e^{-j2\pi f \tau} \\
    & =2N \sum_{\tau = -N+1}^{N-1} \delta[\tau]\, e^{-j2\pi f \tau} = 2N. 
   \end{aligned}
\end{equation}We can see that the power spectra of a Golay complementary sequence pair add up to a constant.

In the same way, we define below a Golay complementary array pair \cite{jedwab2007golay}.

\begin{definition}[Golay complementary array pair]
\label{def:Golay-array_pair}
    Unimodular arrays $\vect{U} \in \mathbb{C}^{N_1\times N_2}$ and $\vect{W} \in \mathbb{C}^{N_1 \times N_2}$ form a Golay complementary array pair if 
    \begin{align}
    \label{eqn:conditionGolay2}
    	R_{\vect{U}}[\tau_1,\tau_2] + R_{\vect{W}}[\tau_1,\tau_2] &= 2N_1 N_2  \delta [\tau_1,\tau_2],
    \end{align}with $R_{\vect{U}} [\tau_1,\tau_2]$ being the ACF of $\vect{U}$ given at the top of the next page.
      \begin{figure*}[t]
    \begin{small}
    \begin{align}
    \label{eq:ACF2}
    	R_{\vect{U}}[\tau_1,\tau_2] = \left\{ 
    	\begin{aligned} 
    	&\sum\limits_{n_1=1}^{N_1-\tau_1} \sum\limits_{n_2=1}^{N_2-\tau_2} [\vect{U}]_{n_1,n_2} [\vect{U}]_{n_1+\tau_1,n_2+\tau_2}^*, 
    	 & & \tau_1=0,\ldots,N_1-1, \tau_2 = 0,\ldots,N_2-1,\\
    	&\sum\limits_{n_1=1}^{N_1+\tau_1} \sum\limits_{n_2=1}^{N_2-\tau_2} [\vect{U}]_{n_1-\tau_1,n_2} [\vect{U}]_{n_1,n_2+\tau_2}^*, 
    	 & & \tau_1=-N_1+1,\ldots,-1, \tau_2 = 0,\ldots,N_2-1,\\
      &\sum\limits_{n_1=1}^{N_1-\tau_1} \sum\limits_{n_2=1}^{N_2+\tau_2} [\vect{U}]_{n_1,n_2-\tau_2} [\vect{U}]_{n_1+\tau_1,n_2}^*, 
    	 & & \tau_1=0,\ldots,N_1-1, \tau_2 = -N_2+1,\ldots,-1,\\
      &\sum\limits_{n_1=1}^{N_1+\tau_1} \sum\limits_{n_2=1}^{N_2+\tau_2} [\vect{U}]_{n_1-\tau_1,n_2-\tau_2} [\vect{U}]_{n_1,n_2}^*, 
    	 & & \tau_1=-N_1+1,\ldots,-1, \tau_2 = -N_2+1,\ldots,-1,\\
    	& \qquad \qquad \qquad 0,
    	 & & \mathrm{otherwise}.
    	\end{aligned} \right. 
    \end{align}
      \end{small}
        \hrulefill
      \end{figure*}

\end{definition} 

Taking similar steps as before, we can show that the sum of the PSDs of a Golay complementary array pair is a constant, i.e.,
\begin{equation}
\label{eq:PSD_2D}
    S_{\vect{U}}(f_1,f_2) + S_{\vect{W}}(f_1,f_2) = 2N_1 N_2.
\end{equation}

\subsection{RIS Configuration Design}
The following proposition shows the connection between Golay complementary array pairs and dual-polarized RIS phase configurations. 

\begin{proposition}
\label{prop:Golay_pair}
Consider a dual-polarized RIS with the phase configuration matrices $\boldsymbol{\Upsilon}_{\mathrm{p}} \in \mathbb{C}^{N_y \times \Tilde{N}_z},\, \mathrm{p} \in \{\mathrm{H},\mathrm{V}\}$, where the first column of $\boldsymbol{\Upsilon}_{\mathrm{p}}$ is formed by the first $N_\mathrm{y}$ entries of $\boldsymbol{\phi}_{\mathrm{p}}$, the second column consists of the second $N_\mathrm{y}$ entries of $\boldsymbol{\phi}_{\mathrm{p}}$ and so on. The RIS radiates a perfectly broad beam if and only if $(\boldsymbol{\Upsilon}_{\mathrm{H}},\boldsymbol{\Upsilon}_{\mathrm{V}})$ form a Golay complementary array pair.

\end{proposition}

\begin{IEEEproof}
  To radiate a perfectly broad beam, the per-polarization configurations of the RIS must satisfy \eqref{eq:PDAF_constant}. Using \eqref{eq:PDAF}, we can rewrite the condition in \eqref{eq:PDAF_constant} as  
\begin{equation}
\begin{aligned}
\label{eq:PDAF_constant_expanded}
 \sum_{\mathrm{p}\in\{\mathrm{H,V}\}} \left| \sum_{n_z=1}^{\tilde{N}_z} \sum_{n_y=1}^{N_y}  [\boldsymbol{\Upsilon}_{\mathrm{p}}]_{n_y,n_z} e^{-j \left((n_y - 1)\hat{\psi}_y + 2(n_z - 1)\hat{\psi}_z \right)}\right|^2 = c
 \end{aligned}
\end{equation}in which  
\begin{equation}
   \hat{\psi}_y = \psi_y + \tilde{\psi}_y,~~\hat{\psi}_z = \psi_z + \Tilde{\psi}_z,
\end{equation}where $\Tilde{\psi}_y$ and $\Tilde{\psi}_z$ are obtained from \eqref{eq:psi} by replacing $(\varphi,\theta)$ with $(\Tilde{\varphi},\Tilde{\theta})$. The terms inside the absolute sign in  \eqref{eq:PDAF_constant_expanded} are the square of the 2D discrete-space Fourier transform of  $\boldsymbol{\Upsilon}_{\mathrm{H}}$ and $\boldsymbol{\Upsilon}_{\mathrm{V}}$. Since the magnitude squared of the Fourier transform of an array equals its PSD  \cite[Chapter 4]{Stein2000}, \eqref{eq:PDAF_constant_expanded} is equivalent to  
\begin{equation}
\label{eq:sum_spectral_density}
 S_{\boldsymbol{\Upsilon}_{\mathrm{H}}}(\hat{\psi}_y,\hat{\psi}_z) + S_{\boldsymbol{\Upsilon}_{\mathrm{V}}}(\hat{\psi}_y,\hat{\psi}_z) = c.
\end{equation} Taking inverse Fourier transform from both sides of \eqref{eq:sum_spectral_density}, we arrive at 
\begin{equation}
\label{eq:sum_ACF_RISconfigs}R_{\boldsymbol{\Upsilon}_{\mathrm{H}}}[\tau_y,\tau_z] + R_{\boldsymbol{\Upsilon}_{\mathrm{V}}}[\tau_y,\tau_z]  = c\delta [\tau_y,\tau_z], 
\end{equation}which signifies that the ACFs of matrices $(\boldsymbol{\Upsilon}_{\mathrm{H}},\boldsymbol{\Upsilon}_{\mathrm{V}})$ must add up to a Kronecker delta function if we want the RIS to radiate a perfectly broad beam (with a spatially flat array factor). Therefore, $(\boldsymbol{\Upsilon}_{\mathrm{H}},\boldsymbol{\Upsilon}_{\mathrm{V}})$ must form a Golay complementary array pair in which case we have $c = 2 \Tilde{N}_z N_y = N_z N_y$. Therefore, if the dual-polarized RIS radiates a broad beam, then, its phase configuration matrices form a Golay complementary array pair. 

On the other hand, if the RIS phase configuration matrices $(\boldsymbol{\Upsilon}_{\mathrm{H}},\boldsymbol{\Upsilon}_{\mathrm{V}})$ constitute a Golay complementary array pair, \eqref{eq:PSD_2D} yields that 
\begin{equation}
\label{eq:sum_PSD}
 S_{\boldsymbol{\Upsilon}_{\mathrm{H}}}(f_1,f_2) + S_{\boldsymbol{\Upsilon}_{\mathrm{V}}}(f_1,f_2) = 2 \Tilde{N}_z N_y = N_z N_y.
\end{equation}Writing the PSDs in \eqref{eq:sum_PSD} as the magnitude squared of the Fourier transform of the matrices, we arrive at 
\begin{equation}
\begin{aligned}
    \label{eq:PDAF_constant_expanded2}
 \sum_{\mathrm{p}\in\{\mathrm{H,V}\}} \left| \sum_{n_z=1}^{\tilde{N}_z} \sum_{n_y=1}^{N_y}  [\boldsymbol{\Upsilon}_{\mathrm{p}}]_{n_y,n_z} e^{-j \left(n_y f_1 + n_z f_2 \right)}\right|^2 =N_z N_y.
 \end{aligned}
\end{equation}Setting $f_1 = \hat{\psi}_y $ and $f_2 = 2\hat{\psi}_z$, and multiplying the terms inside the absolute sign by $e^{j(\hat{\psi}_y + \hat{\psi}_z)}$, we end up at 
\begin{equation}
   A(\varphi,\theta) = N_z N_y, 
\end{equation}which indicates that an RIS whose configuration matrices form a Golay array pair produces a perfectly broad beam with a spatially flat power-domain array factor. The proof is thus completed. 
\end{IEEEproof}

\section{Construction of Golay Complementary Array Pairs for Broad Dual-Polarized Beamforming}
Many Golay complementary sequence pairs have been identified via exhaustive numerical search, though it is known that for some sequence lengths, no Golay sequence pair exists in the quaternary alphabet \cite{Holzmann94}. Here, we describe how to construct Golay complementary array pairs based on known Golay complementary sequence pairs.

\begin{proposition}
    \label{prop:Golay-expansion}
Let $(\vect{u}_1,\vect{w}_1)$ and $(\vect{u}_2,\vect{w}_2)$ be two pairs of Golay complementary sequences of lengths $L_1$ and $L_2$, respectively. Then, $(\vect{U},\vect{W})$ formed by either
\begin{equation}
\begin{aligned}
    	\label{eq:Golay-expansion1}
    	&\vect{U} = \begin{bmatrix}
    		\vect{u}_1 \vect{u}_2^{\Ttran}\\
    		-\vect{w}_1\vect{w}_2^{\Htran}\vect{E}_{L_2}
    	\end{bmatrix} \in \mathbb{C}^{2L_1 \times L_2},\\
    &\vect{W} = \begin{bmatrix}
    \vect{u}_1\vect{w}_2^{\Ttran}\\
    		\vect{w}_1 \vect{u}_2^{\Htran}\vect{E}_{L_2}
    	\end{bmatrix}\in \mathbb{C}^{2L_1 \times L_2},
    \end{aligned}
    \end{equation}or
    \begin{equation}
        \begin{aligned}
        \label{eq:Golay-expansion2}
       &\vect{U} = \left[\vect{u}_1 \vect{u}_2^{\Ttran},\; -\vect{w}_1\vect{w}_2^{\Htran}\vect{E}_{L_2}\right] \in \mathbb{C}^{L_1 \times 2L_2}, \\
       & \vect{W} = \left[\vect{u}_1\vect{w}_2^{\Ttran},\; \vect{w}_1 \vect{u}_2^{\Htran}\vect{E}_{L_2}\right]\in \mathbb{C}^{L_1 \times 2L_2}
        \end{aligned}
    \end{equation}
is a Golay complementary array pair, where $\vect{E}_L$ is an $L \times L$ matrix having ones on the anti-diagonal and zeros elsewhere.  
    	
\end{proposition}
\begin{IEEEproof}
    The proof will be provided in the longer version of this paper.
\end{IEEEproof}
\section{Numerical Results}
\begin{figure*}
	\centering
	\begin{subfigure}{0.33\textwidth}
		\centering
		\includegraphics[width=\columnwidth]{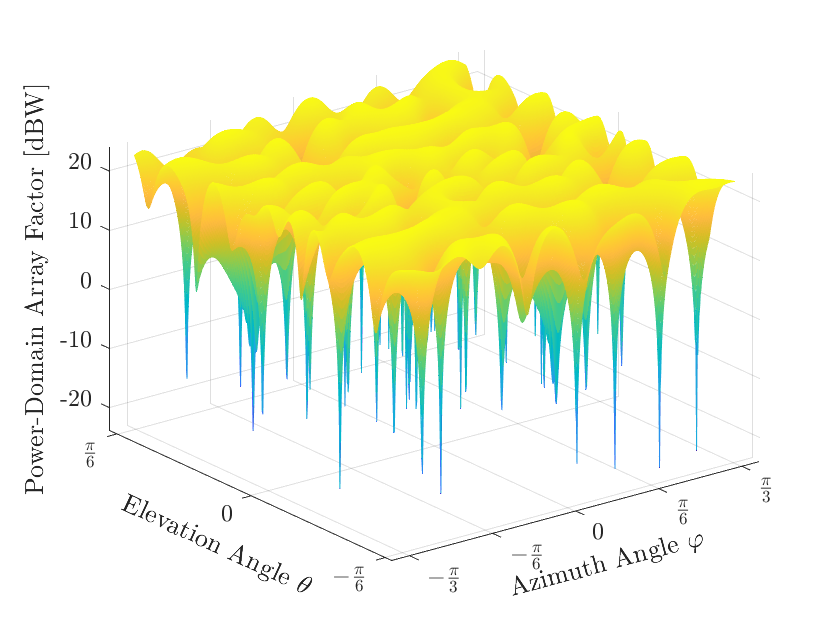}
		\caption{Polarization H}
		\label{fig:AF-polH}
	\end{subfigure}%
	\begin{subfigure}{0.33\textwidth}
		\centering
		\includegraphics[width=\columnwidth]{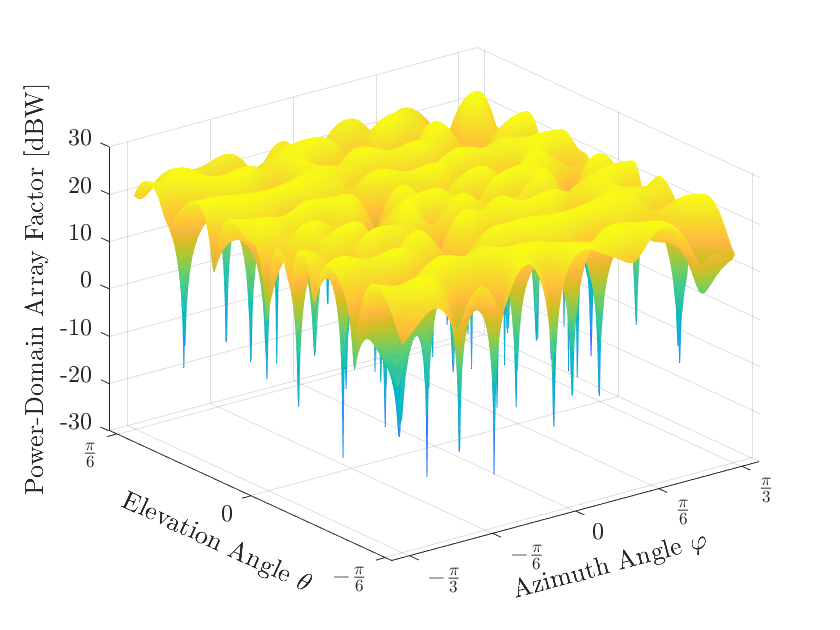}
		\caption{Polarization V}
		\label{fig:AF-polV}
	\end{subfigure}%
 \begin{subfigure}{0.33\textwidth}
		\centering
		\includegraphics[width=\columnwidth]{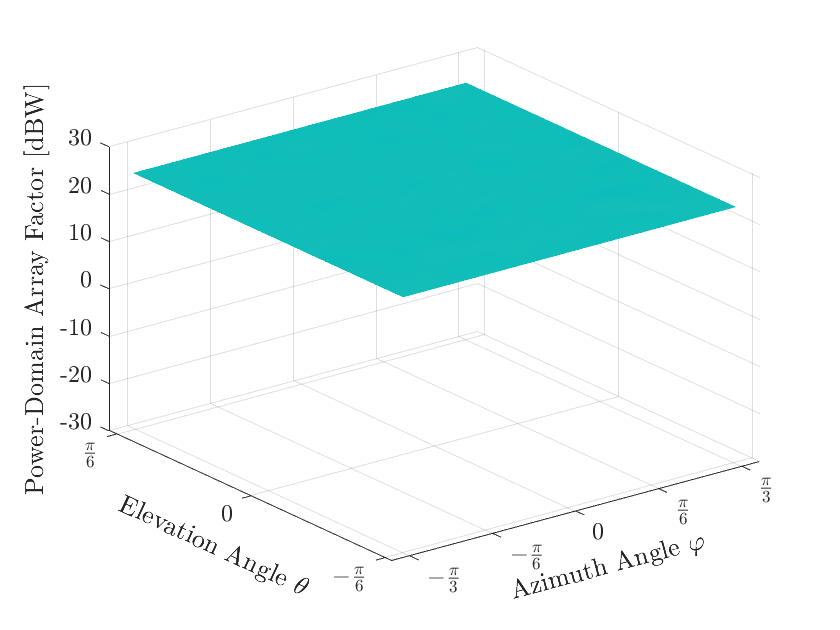}
		\caption{Total}
		\label{fig:AF-tot}
	\end{subfigure}
	\caption{Illustration of the per-polarization and total power-domain array factor for a dual-polarized RIS for which the phase configuration matrices of the two polarizations form a Golay complementary array pair. The total gain is flat while the individual polarizations have peaks and valleys. The considered RIS has $256$ elements with $N_y = N_z = 16$ and an inter-element spacing of $\Delta_y = \Delta_z = \lambda/2$.}
 \label{fig:AF}
\end{figure*}
Herein, we provide numerical simulations to corroborate the theoretical analyses presented earlier. We consider an RIS with $N_y = N_z = 16$ and $\Delta_y = \Delta_z = \lambda/2$. The RIS assists the downlink broadcast communication between a BS and a number of users uniformly distributed around the RIS where the AoDs from the RIS to the users are given by $\varphi \in [-\pi/3,\pi/3]$ and $\theta \in [-\pi/6,\pi/6]$. Moreover, the AoAs to the RIS are set as $\Tilde{\varphi} = -\pi/3$ and $\Tilde{\theta} = \pi/3$. In the RIS setup described above, each polarization consists of $128$ elements in the form of a $16 \times 8$ UPA. We use Proposition~\ref{prop:Golay-expansion} to obtain the phase configuration matrices of the RIS using two known Golay complementary sequence pairs of length $8$. Specifically, we utilize the following two pairs of Golay sequences \cite{Holzmann94} for the construction of a Golay complementary array pair:
\begin{equation}
    \begin{aligned}
     &\vect{u}_1 = \exp \left(j\left[0,0,0,0,0,0,\pi,\pi,0\right]^{\Ttran}\right),\\
     &\vect{w}_1 = \exp \left(j\left[0,0,\pi,\pi,0,\pi,0,\pi\right]^{\Ttran}\right),
    \end{aligned}
\end{equation}and
\begin{equation}
    \begin{aligned}
     &\vect{u}_2 = \exp \left(j\left[0,0,0,0,0,\frac{\pi}{2},\frac{-\pi}{2},\frac{-\pi}{2},\frac{\pi}{2}\right]^{\Ttran}\right),\\
     &\vect{w}_2 = \exp \left(j\left[0,0,\pi,\pi,\frac{\pi}{2},\frac{-\pi}{2},\frac{\pi}{2},\frac{-\pi}{2}\right]^{\Ttran}\right).
    \end{aligned}
\end{equation}Then, the two phase configuration matrices of the RIS for polarizations $\mathrm{H}$ and $\mathrm{V}$ are formed by \eqref{eq:Golay-expansion1}.

Fig.~\ref{fig:AF} illustrates the power-domain array factor for the considered RIS, where the array factors for polarizations $\mathrm{H}$ and $\mathrm{V}$ are given in Fig.~\ref{fig:AF-polH} and Fig.~\ref{fig:AF-polV}, respectively. Fig.~\ref{fig:AF-tot} shows the total power-domain array factor obtained by summing up the power-domain array factors of the two polarizations. It can be observed that the total power-domain array factor is spatially flat over the angular area of interest, verifying the theoretical result given in Proposition~\ref{prop:Golay_pair}, while the individual polarizations have both peaks and valleys.
The constant total power-domain array factor is  $A(\varphi,\theta) = 10\log(N_y N_z) = 24.08\,$dBW for all the considered azimuth and elevation angles. 
\begin{figure}
    \centering
    \includegraphics[width=0.8\columnwidth]{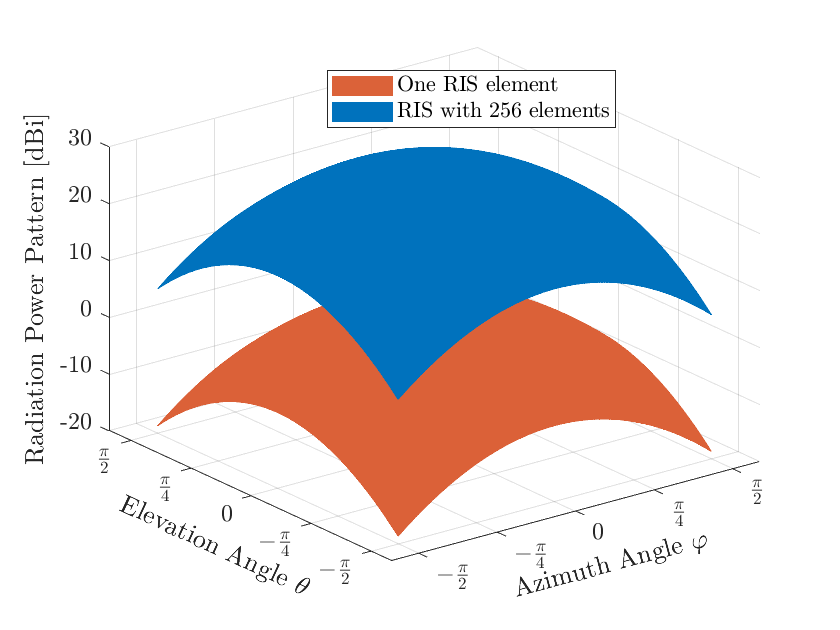}
    \caption{Illustration of the radiation power pattern for a $256$-element dual-polarized RIS when its phase configuration matrices are a Golay complementary array pair.  }
    \label{fig:radiation-pattern}
\end{figure}

Fig.~\ref{fig:radiation-pattern} shows the total radiation power pattern of the RIS over $\varphi \in [-\pi/2,\pi/2]$ and $\theta \in [-\pi/2, \pi/2]$. The radiation pattern of one RIS element is modeled by the 3GPP antenna gain model \cite{3gpp},
\begin{align}
    G_{R,0}(\varphi,\theta) =&\; 8 -\min \bigg[ \min \Big[12 \left( \frac{\varphi - \varphi_0}{\Delta \varphi}\right)^2,30 \Big] \notag \\ & + \min \Big[12 \left( \frac{\theta - \theta_0}{\Delta \theta}\right)^2,30 \Big], 30 \bigg]~~~\mathrm{[dBi]},
\end{align}with $\varphi_0 = \theta_0 = 0$ and $\Delta \varphi = \Delta \theta = \pi/2$. Considering the two-hop transmission from the BS to the RIS and from the RIS to the users, the total radiation power pattern of one RIS element would be obtained as $G_{R,0}(\Tilde{\varphi},\Tilde{\theta})G_{R,0}(\varphi,\theta)$. The figure shows that the RIS whose configuration matrices form a Golay complementary array pair preserves the broad radiation pattern of a single RIS element, while shifting it upward. This, once again, endorses the effectiveness of the proposed approach for producing a broad reflected beam from the RIS. 

\section{Conclusions}
To have a perfectly broad reflected beam from an RIS, the power-domain array factor of the surface must be spatially flat. This is not achievable with a uni-polarized RIS because the breadth of the radiation pattern is inversely proportional to the aperture size of the RIS which is typically large. However, we can achieve uniform broad beamforming if a dual-polarized RIS is utilized. The key is to design the RIS configurations of the two polarizations such that the beams radiated from different polarizations complement each other and create an overall broad radiation pattern. We showed both analytically and numerically that this can be accomplished by setting the RIS configuration matrices as a Golay complementary array pair. Since the sum of the ACFs of a Golay complementary pair adds up to a Kronecker delta, an RIS configured with a Golay pair yields a spatially flat power-domain array factor.

\bibliographystyle{IEEEtran}
\bibliography{refs}
\end{document}